# Return to Lacan: an approach to digital twin mind with free energy principle

Lingyu Li [1,2], Chunbo Li [*,1,2]

[1] Shanghai Mental Health Center, Shanghai Jiao Tong University School of Medicine, Shanghai 200030, China; [2] Shanghai Jiao Tong University School of Medicine, Shanghai 200023, China

**Abstract**

Free energy principle (FEP) is a burgeoning theory in theoretical neuroscience that provides a universal law for modelling living systems of any scale. Expecting a digital twin mind from this first principle, we propose a macro-level interpretation that bridge neuroscience and psychoanalysis through the lens of *computational Lacanian psychoanalysis*. In this article, we claim three fundamental parallels between FEP and Lacanian psychoanalysis, and suggest a FEP approach to formalizing Lacan's theory. Sharing the non-linear temporal structure that combines prediction and retrospection (*logical time*), both of two theories focus on epistemological questions that how systems represented themselves and external world, and those elements failed to be represented (*lacks* and free energy) significantly influence the systems' subsequent states. Additionally, the fundamental hypothesis of FEP that the precise state of environment is always concealed, accounts for *object petit a*, the core concept in Lacan's theory. With neuropsychoanalytic mapping from three orders (the Real, the Symbolic, and the Imaginary, RSI) onto brain regions, we propose a brain-wide FEP model for a minimal definition of Lacanian mind – composite state of RSI that is perturbated by desire running over the logical time. The FEP-RSI model involves three FEP units connected by respective free energy with a natural compliance with logical time, mimicking core dynamics of Lacanian mind. The biological plausibility of current model is considered from perspectives of cognitive neuroscience. In conclusion, the FEP-RSI model encapsulates a unified framework for digital twin modeling at the macro level.

**Introduction**

Cognitive computational neuroscience proposes an ambitious integration among computational neuroscience, cognitive science, and artificial intelligence, with the goal of explaining neuronal activities and cognitive functions utilizing biological plausible computational models (Kriegeskorte & Douglas, 2018). Computational models formally define the process of brain representation, by solving which on computer, corresponding process can be imitated, called computational simulation (Durán, 2020). Studies in this field have proposed various computational model of specific cognitions ranging from perception, object recognition, action, and memory to language, social cognition, and consciousness etc. An appealing idea which naturally emerges but remains elusive is to encapsulate existing models of these intertwined cognitive functions and hence assemble a digital twin. By mirroring physical life, digital twin can help tracking health condition, detecting and repairing potential problems, and predicting future status (Tao et al., 2022). However, intruding challenges in a digital twin for human-mind rooted in its nature as a complex system, including but not limited in heterogeneity of multiple modalities in their functions, structures, scales, and hierarchies, unknown connectivity patterns among various cognitions, and the lack of interpretation theory.

Free energy principle (FEP), a burgeoning theory in theoretical neuroscience, provides a general and formal solution for modeling living systems at any scale ranging from neurons, brain regions, cognitive functions, and individual and collective behaviors (Isomura et al., 2023; Piekarski, 2023). Leveraging FEP, current work formalized Lacanian psychoanalysis, hence offering a potential approach to digital twin mind at the micro level. Lacan's theory represents an extremely complex interpretation of psychic dynamics, which are treated as a composite of three orders – the Real, the Symbolic, and the Imaginary (Evans, 2006). These three orders cover most cognitive modalities, depicting the perceptive and linguistic representations of self and the world. Based on the fundamental parallels underlying the two theories, this article introduced computational Lacanian psychoanalysis and a roadmap toward a digital twin at macro level.

**Lacanian psychoanalysis**

Lacan is renowned for his sophisticated theory of psychoanalysis, deeply rooted in philosophy and linguistics. At the core of Lacanian psychoanalysis are the three orders: the Real, the Symbolic, and the Imaginary (RSI). These orders serve as a fundamental framework for understanding where the human subject resides within their mental states. According to Lacan, an individual's mental state is a blend of these three domains that influence the subject simultaneously and interdependently, akin to a Borromean ring. The Imaginary represents a perceptive and internal representation of the external world. The dynamic interplay between the external world and the internal self leads to primary self-identification and self-knowledge, though inherently distorted due to its 'imaginary' nature. Consequently, the identification within the Imaginary is essentially a misidentification, forming the basis for psychosis (Mills & Downing, 2018). The Symbolic encompasses the language. Put succinctly, since the "unconscious is structured like a language," the Symbolic operates akin to Hermeneutics, addressing issues of meaning generation, self-interpretation, and intersubjectivity. Hence the Symbolic is linguistic representation of the situations of subject. However, it's important to note that any representation is inherently incomplete, leaving aspects unrepresented. The Real collects the unrepresented, thus existing as a realm of 'impossibility', a missing reality. Derived from the Real, the concept of *repetition*—manifested as an incessant attempt to represent the unrepresented—becomes another fundamental concept of Lacanian psychoanalysis (Feldstein et al., 1995).

The temporal structure of those representations is *logical time*, a notion that human experiences cannot be neatly confined within a unidirectional and chronological understanding of time. From the perspective of logical time, "the past anticipates a future within which it can retroactively find a place". In other words, the past bestows significance upon forthcoming events in an anticipatory manner, and the future imbues the past with retroactive meaning. Along logical time, meaningful relationships between events emerge, transforming time into a mechanism that generates significance. This concept underscores the anticipatory nature of human mind and emphasizes the importance of retroactive reconstruction (Hook et al., 2022).

And what drives the ongoing evolution of the three orders within logical time? Repetition, as mentioned, serves as the driving force. For Lacan, repetition represents the return of something that maintains identical—*object petit a*, a concept rooted in the Real, leading to the endless metonymic course of desire. And orientation of *drive* is the running of desire towards object petit a. The history of subjectivity unfurls in the repetition of desire's trajectory: the unrepresented → object petit a → desire → object of desire → failure of representation → the unrepresented. That is, desire can never be satisfied, and the endless cycles contribute to so-called *fantasy* that need to be traversed by psychoanalysis practice. A minimal definition of the human mind from Lacanian perspective ultimately emerges — composite status of three orders which is perturbated by desire running over the logical time.

**Free energy principle**

To survive, living systems must interact with environment, including precepting and changing it. Free energy principle (FEP) provides a universal principle for modeling these interactions between environment and organisms at any scales like neuron, brain region, and individual and collective behaviors (Isomura et al., 2023; Piekarski, 2023). Similar to Kant's notion that we can only perceive phenomena and not the thing-in-itself, the starting point of FEP is that precise state of the environment is usually concealed (referred to as the *hidden state*), and living systems thus can only infer the hidden state through observations. The entity implementing inferences is defined as *internal model* within systems, and FEP formalizes the mechanisms of internal models.

In classical Bayesian theory of perception, the belief of the hidden state is defined as the posterior belief according to the Bayes theorem:

$$P(s|o) = \frac{P(o|s)P(s)}{P(o)}$$

For each observation $o$, the internal model infers the hidden state of environment according to existing knowledge on the hidden state, known as *priors* $P(s)$, the likelihood of this observation $P(o|s)$, and the overall (marginal) probability of its occurrence $P(o)$. This Bayesian account successfully explains various psychological domains like perceptual constancies, perceptual illusions, cue combination, and multi-sensory integration etc. (Rescorla, 2021). However, the computation of marginal probability $P(o)$ is intractable in complex situation, which significantly hinders the computation of accurate posterior and hence the biological plausibility of basic Bayesian cognitive science. To reach a more efficient and plausible computation, FEP reforms it by involving active predictions and variational inference.

Before precepting, the internal model keeps generating predictions of the forthcoming phenomena $P(o|s)$, based on priors $P(s)$. Therefore, the internal model is also referred as a *generative model*. Now, the living systems are no longer passively inferring the environment, but actively mirroring the world with its internal model, and predicting what will happen next. Therefore, the Bayesian inference is turned into *active inference*. Additionally, the goal of living systems is no longer to compute the accurate posterior according to the Bayes theorem, but to optimize the internal model which can fit the real world as well as possible (Figure 1). This optimization process is formalized by *variational inference*, an approximate Bayesian inference from machine learning (Blei et al., 2017). In the context of variational inference, the term under optimization is called evidence lower bound (ELBO), and in FEP, *variational free energy* equals to the negative ELBO (i.e., maximization of ELBO is equivalent to minimization of variational free energy).

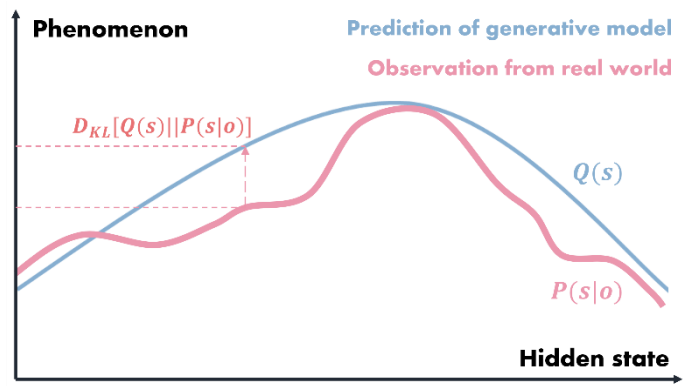

Figure 1. Active inference through generative model mirroring the real world.

Originally a physics concept, *free energy* quantifies the energy available in a thermodynamic system to alter its properties, and system with minimal free energy implies a state of equilibrium. In the context of active inference, living systems with generative models could get rid of repetitive Bayesian inference on exact posterior, but just focus on if the predictions match observations. Only those mismatches are fed back to the system and draw attention. In other words, mismatches induce the non-equilibrium state of generative model, which can be quantified by its variational free energy, gauging the 'energy' required to optimize it. Therefore, an optimal generative model that fits the environment best minimizes the free energy. FEP depicts a same process as the second law of thermodynamic defines – spontaneous process decreases the free energy of the system. Hence, free energy principle can be seen as the physics of mind, both in its form and content (Schoeller, 2019).

Unlike machine learning models of variational inference that merely fit the potential distribution, living systems could change the environment and collect evidence by actions. Therefore, FEP further expands the context to action planning or decision making. That is, the distribution of real world can be changed to fit the internal model. The desired distribution of hidden state is defined as the preferences $C$. For every conceivable action $\pi$, the anticipated state of the system $\tilde{s}$ is the conditional possibilities $Q(\tilde{s}|\pi)$. Again, the best action will render the state of environment fitting preferences. The term under optimization in decision making is the *expected free energy*. In essence, expected free energy is a "belief that one will minimize free energy in the future" (Parr et al., 2022).

**From *object petit a* to free energy**

Basically, we could summarize several consistent philosophies between Lacanian psychoanalysis and FEP. First of all, both theories' scopes focus on epistemological questions: how knowledges or beliefs are acquired and how they are validated. By hypothesizing an internal model that continuously generating simulations and predictions of the world, FEP caters to the modernist transcendental constitution of the experiential world which is created by human rather than found (Sass, 2015). A resembled simulation process also exists in Lacan's theory in the form of psychic reality and physical reality. All perceptions, actions, and beliefs of human are referred to subjectivity – a psychic reality that separates subject and environment (Mooij, 2012). Psychic reality generates expectations, intentions, and 'fantasies', which are constrained by physical reality, and mismatches between psychic and physical reality causing perturbation in the Real order. Mild mismatch induces emotional unpleasure, while severe mismatch could be encoded as traumatic memory in brain, limbic systems and insula especially (Dall'Aglio, 2019; Solms, 2019). From the view of FEP, these two reactions can be seen as the performance and learning of internal models.

Temporal structures of both theories are combination of prediction and post-diction. Interestingly, the two notions are literally aligned: "belief that one will minimize free energy in the future (Parr et al., 2022)" & "the past anticipates a future within which it can retroactively find a place (Hook et al., 2022)". This non-linearly temporal structure is biologically plausible both in real time and long scale. In real time perception, neural systems approximate the present using predictions, and prediction errors are propagated back to correct and overwrite original representations (Hogendoorn, 2022). At a longer time scale, such a temporal structure contributes to sense of agent, conscious intention, and counter-factual inference and so on (Miyamoto et al., 2023; Riemer, 2018).

Another significant similarity is the central role of 'prediction errors', those failed to be represented by internal models. FEP describes living system's behavior as an optimization process to reach the precise simulation on external world by generative model. And free energy, the distance of two distributions, drives the system to learning and acting, i.e., change the distributions of internal model or environment. In this sense, researchers are increasingly aware of that free energy accounts for the psychoanalytic concept *drives* (Sikora, 2022). At the mind level, free energy is perceived as the feeling of unpleasure and anxiety (Albarracin et al., 2024; Solms, 2019), and hence a reverse occurred. The internal model which constitutes the subject becomes unconscious, and only free energy, the failure of predictions, are detected. The first principle of Descartes' philosophy, "I think, there I am", is rewritten by Lacan as "I think where I am not, therefore I am where I do not think" (Lacan, 2001), revealing a totally consistent meaning. The failure of representation is felt as *lack* or *remainder*, arouses drive to solve this lack, and the orientation of this drive is *object petit a*. As claimed by Lacan, object petit a is completely unavailable due to the fundamental split between psychic reality and physical reality. Consequently, subject can only find alternatives of it, i.e., object of desire, again and again, to fill lacks while generating lacks at the same time. From the FEP, this situation can be described as the impossibility of complete fitting between distributions of internal model and the world, which returns to the basic Kantian hypothesis of FEP – the precise state of environment is always concealed.

With these basic parallels between two theories, we believe that a formalized Lacanian psychoanalysis through FEP could efficiently simulates dynamics of subjectivity in depth. Although Lacan's theory is more complicated than that we introduced by now, our grasps of the essence make the rest part as engineering questions.

**Computational Lacanian psychoanalysis: one, two, three, and more**

Equipped with the intuitive understanding of two theories and their similarities, this section aims to propose a mechanistic model for the aforementioned minimal definition of subject – composite status of three orders which is perturbated by desire running over the logical time. Firstly, we propose a brain-wide model of individual subject that consists of three orders. Then the interactive dynamics of two such models are explored to mimic the 'desire running' between subjects. Finally, we expand the context into multi-agent scenario, investigating how collective dynamics of subjects contribute to the culture (or the Lacanian term, the Other) and constrains individual's behaviors. Through the computational modeling and simulation, we demonstrate the potential of computational Lacanian psychoanalysis to mimicking psychic dynamics and interpreting digital twin mind at the macro level. Original code is available in GitHub (https://github.com/YabYum/ActiveInferenceLacan).

*One: subject as a Borromean knot*

First of all, we begin with an initial endeavor to dynamically map the functions of the three orders onto the brain (Figure 2. a), setting the stage for our formal model. Our intention is not to achieve a precise anatomical or functional mapping, but rather to establish an intuitive framework for defining the system of current interest. The Real is situated within the upper brainstem and diencephalic system, as these areas play a fundamental role in affective experiences, consciousness, and the primary needs of the body such as sustenance, sexuality, and homeostasis (Dall'Aglio, 2019). The Imaginary corresponds to the parietal and occipital lobes, given their involvement in motor control, visual perception, and the representation of self (body-image). The Symbolic domain is allocated to the prefrontal and parietal lobes, responsible for executing language processing, generating meaning, and conducting thought experiments (Dall'Aglio, 2019; Miyamoto et al., 2023). Numerous studies have leveraged FEP to model various cognitive functions across the brain, as summarized in table 1. A hierarchical model that encompasses these detailed elements corresponding to three orders can fulfill a more refined digital twin, but we will simplify this model and focus on an intuitive interpretation at the top-tier hierarchy initially. Also, we assume that three orders are ontologically equal entities that share FEP as the first principle.

**Table 1. Existing studies modeling cognitive functions using free energy principle**

| Order | Sub-element | Representative paper |
|---|---|---|
| *The Real* | Emotion | (Smith et al., 2019) |
| | Interoception | (Paulus et al., 2019) |
| | Consciousness | (Vilas et al., 2022) |
| | Physiology | (Sedley et al., 2024) |
| *The Symbolic* | Communication | (Friston & Frith, 2015) |
| | Self-esteem | (Albarracin et al., 2024) |
| | Culture | (Kastel et al., 2023a) |
| | Understanding | (Parr & Pezzulo, 2021) |
| *The Imaginary* | Self/other distinction | (Lanillos & Cheng, 2020) |
| | Intentional actions | (Priorelli & Stoianov, 2023) |
| | Self-image | (Tremblay et al., 2021) |
| | Theory of mind | (Hipólito & van Es, 2022) |

Every order is assigned with an isomorphic generative model (Figure 2. b). Zooming into the basic unit (Figure 2. c), at time step $\tau$, the unit has a prior belief on current state $P(s)$ and predicts potential observations based on likelihood $P(o|s)$. After receiving real observation, the approximate posterior of hidden state is updated as:

$$Q(s) = \ln P(o|s) + \ln P(s)$$

Variational free energy is as follows:

$$F[Q, o] = D_{KL}[Q(s)|P(s)] - \mathbb{E}_{Q(s)}[\ln P(o|s)]$$

The first term on right hand side is the KL divergence between current posteriors and priors, and the second term is negative log of predictions, i.e., model evidence. To reduce variational free energy or realize prefer states C with actions, transition probability is necessary that evaluate the result of specific policy $\pi$ — potential state caused by this action $Q(\tilde{s}|\pi)$.

Corresponding potential observations is predicted via likelihood $P(\tilde{o}|\tilde{s})$. Then the expected free energy of candidate policies is calculated:

$$G(\pi) = \mathbb{E}_{Q(\tilde{s}|\pi)}[H[P(\tilde{o}|\tilde{s})]] + D_{KL}[Q(\tilde{s}|\pi)||P(\tilde{s}|C)]$$

The policy with minimal expected free energy is believed to induce minimal variational free energy when taken.

Message passing and belief updating processes are applied to connect three function units (Champion et al., 2021). Variational message passing is traditionally utilized for hierarchical generative models that implement high-order representation from low-order representations, for example, connectivity pattern of six-layers neocortex. However, we assume a connectivity between different cognitive modalities, which is recurrent rather than hierarchical. To be specific, variational free energy emerging from each order propagates globally to other units and hence connects three units together with variable weights (Figure 2. b). Intuitively, errors in one cognition can influence performances of another one. For instance, chronic social stressors like unemployment cause abnormal physiological activities (Jandackova et al., 2012). Lacan describes three intertwined orders as a Borromean knot, and the *object petit a* is the node that connects them, and the connectivity represents dynamics of drive among three orders. Additionally from neuroscientific lens, dopamine could generally encodes prediction errors and transmits across brain (Gardner et al., 2018), ensuring the biological plausibility of such a connectivity pattern.

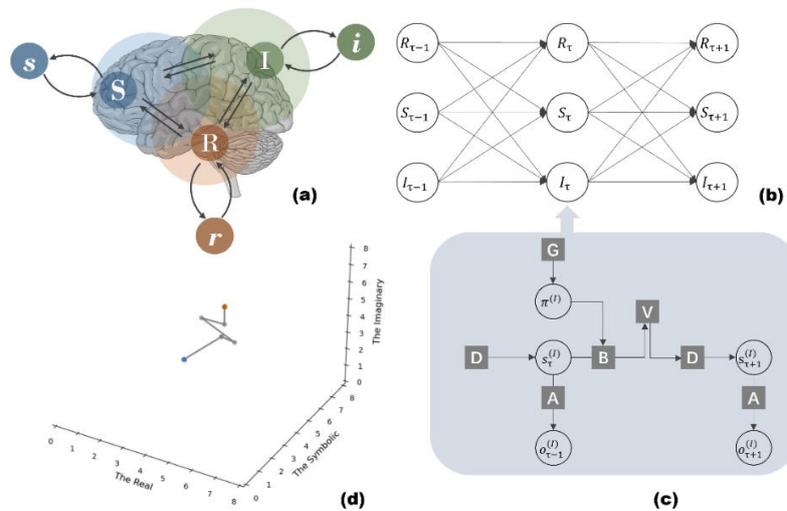

Figure 2. Illustration of recurrent generative model entailing principles of Lacanian psychoanalysis and active inference. (a) An intuitive mapping of three orders onto brain regions. Capital and lowercase letters denote

**generative model and corresponding environment, respectively. R: the Real; S: the Symbolic; I: the Imaginary. (b) Flowchart of our recurrent generative model with three interconnected basic units operating in discrete time. (c) A closer look at basic unit within the generative model. (d) Simulation of dynamics of three orders when the Symbolic order is perturbated.**

Through this recurrent model, an integrative psyche could be modelled, involving multiple interconnected cognitive components. For example, we can reproduce the dynamics of depressed mood by manually inducing free energy in the Symbolic order, which represents life events against expectations. Therefore, we simulate the 15-timesteps dynamics of the three orders when the Symbolic order is perturbed, utilizing Python 3.10 (Heins et al., 2022). We set the initial state (i.e., priors) of the Symbolic order at 0, with a preference value of 4, while maintaining consistent priors and preferences for the other two orders. As depicted in Figure 2.d, when the Symbolic order aligns with its preferred position, the other two orders exhibit synchronous fluctuations, representing the interconnectivity between three orders. This interconnectivity stems from the brain-wide propagation of divergence item and residual free energy. The simulated dynamics can be interpreted as that variational free energy raising from life events against expectations, is propagated to other two domains and hence disturbs their stable states, inducing various psychopathological symptoms like emotional responses, autonomic nerves dysfunction, disturbed body perception, and so on. For a Lacanian interpretation, malfunction within the Symbolic order, leading to lacks in the Other, causing the return to *object petit a*. At the same time, his lack disrupts the typical mediatory function of the Symbolic, preventing the subject from adequately symbolizing experiences rooted in the Real, therefore induces significant distress in the Imaginary order (Hook & Vanheule, 2022).

*Two: desire as partial generalized synchronization*

In this section, we aim to integrate the concept of desire into our model to capture the dynamics of the psyche within interpersonal contexts. In the field of FEP, interpersonal dynamics, communications, are studied with the paradigm involving two subjects with similar internal models (Friston & Frith, 2015; Pan et al., 2023; Vasil et al., 2020). To realize an effective communication, two agents need to infer each other's state and predict behaviors based on their models that hence reach a synchronization – referred to as *synchronization of chaos* or *generalized synchronization*.

Lacan frames the essence of desire as *metaphor*, in a linguistic fashion. That is, envisioning two subjects as signifiers within a signifying chain, desire of subject is to seizure the signified of an object (Fink, 2017). From the lens of complexity theory, the desire of one subject on other manifests as a tendency toward a generalized synchronization of their Symbolic orders. Since that our model consists of three orders, the process of desire, synchronization of the Symbolic orders, entails *partial generalized synchronization* (Figure 3. **a**). This partial synchronization enables that two synchronized subjects maintain their respective chaotic behaviors (Hasler et al., 1998). However, it is worth noting that the direct synchronization of the Symbolic order is impossible because of the barriers of language as Lacan argued in his L-scheme, and thus the partial synchronization here is fantasies of subjects (Lacan, 1988).

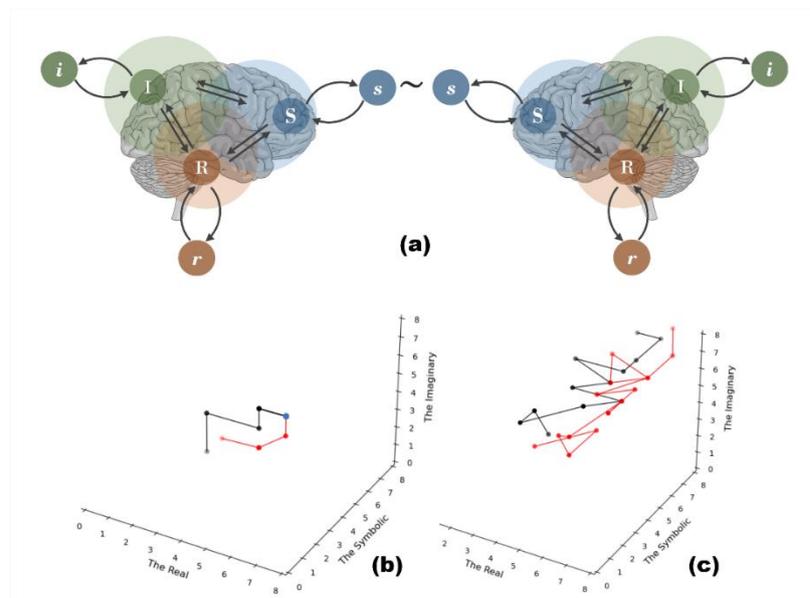

**Figure 3. Desire as partial generalized synchronization between individuals. (a) Two agents, with the same internal model but different initial states, infer each other's hidden state of the Symbolic orders as their own preferences, which guarantee a partial generalize synchronization. When simulating such dynamics, we found that under identical condition, subsequent dynamics could be highly variable, and we display two of them, as illustrated in (b) and (c).**

To illustrate such a partial generalized synchronization, we utilize the recurrent model proposed in last section, and employ two subjects (A & B) with shared FEP-RSI internal model but differing initial states. Subjects A and B keep inferring the state of each other's

Symbolic order, and treating each other's states as their own preferences. The dynamics of these two subjects' partial synchronization are illustrated in Figure 3. **b** and **c**, spanning 15-timesteps. As the Figure 3. **b** shows, stemming from the interconnective nature of three orders, an ideal partial synchronization of their Symbolic orders can indirectly lead to the convergence of the other two orders. The perfect unification of two subjects reveals ultimate goal of desire – a complete fulfillment of object petit a where we find Lacan's formula of desire:

$$\$ \diamond a$$

which means desire is the relationship between (barred) subject and object petit a. The perfect synchronization can only exist before one's birth, and once born, subject is plunged into a world of others. The first attempt of synchronization with mOther demonstrates the Oedipus complex of which solution is to enter the Symbolic order through internalizing the language. As the psychic structure increasingly complex, a sense of total convergence and unity, minimization of object petit a, like in the uterus, becomes more and more impossible (Ormrod, 2014). Despite the minimal complexity of our model, the subsequent dynamics exhibit significant diversity under identical initial condition. In Figure 3. **b**, the two subjects achieve rapid synchronization, whereas in Figure 3. **c**, the synchronization remains partial. This further reflects the multifaceted and random nature of human interaction and communication. Another interesting observation is the indirect convergence through partial generalized synchronization, this dynamic aligns with studies indicating that romantic partners often exhibit shared autonomic physiology and emotional regulation across various timeframes (Butler & Randall, 2013; Ogolsky et al., 2022; Palumbo et al., 2017). One plausible interpretation is that the internal models of romantic partners tend to align over time due to the effects of long-term synchronization. In summary, by unveiling desire as a mechanism driving partial synchronization between individuals, we formalized the mechanisms of desire running, successfully expanding the FEP-RSI model from single subject to interpersonal situations.

*Three and more: the Other as collective dynamics*

Involving multiple agents into the current context, FEP has been applied into collective behaviors and cumulative culture (Kastel & Hesp, 2021; Kastel et al., 2023b). With each participant acting on the principle of minimizing free energy, diverse collective dynamics could emerge spontaneously without explicitly setting any rule (Heins et al., 2023). Remarkably, participants in the self-organized system percept the environment, infer their position, and subsequently become functionally segregated. As articulated by researchers, these participants "infer what specific functional role they play based on the early inputs they receive" (Ramstead et al., 2021). This idea naturally leads us to the big Other (or the Other with a capital O) in the Lacanian psychoanalysis, which manifests as culture, language, society, ideology, and collective dynamics (Evans, 2006).

Before discussing the vast concept, we can firstly examine the dynamics of a minimal group. We design a partial generalized synchronization process similar to that in last section but involving three subjects with shared FEP-RSI model. The synchronizing goal is a triangle – A desires B, B desires C, and C desires A, as the Figure 4. **a** depicted. The simulation in Figure 4. **b** displays the trio, featuring rich divergences and convergences. Interestingly, by desiring B who desires C, A could sometimes indirectly synchronize with C, and vice versa. It may be the clearest illustration of Lacan's famous assertion "desire is the desire of others' desire" (Lacan, 2001). And expanding the trio to collective dynamics composed by the whole society, man's desire naturally becomes "the desire of the Other".

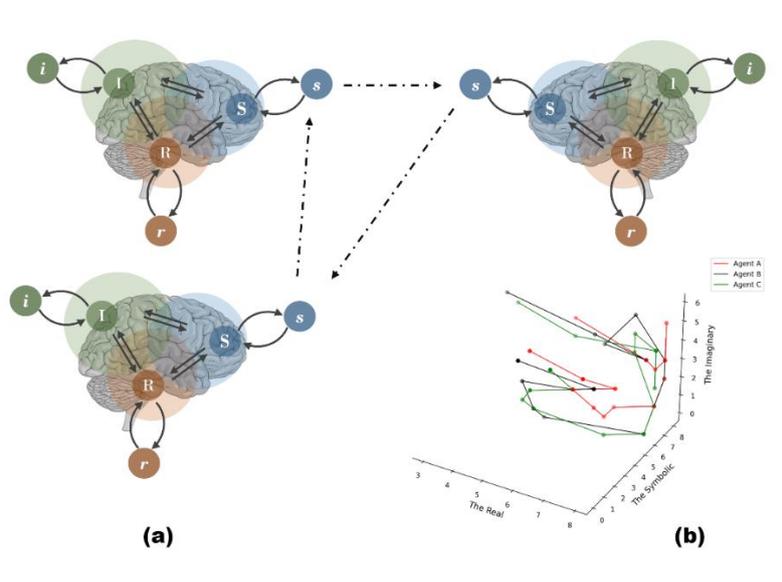

**Figure 4. Collective dynamics of a minimal group as partial generalized synchronization between three individuals. (a) Three agents, with the same internal model but different initial states, implement partial**

**generalize synchronization of their Symbolic orders in a triangle structure. (b) simulations of such collective dynamics.**

The Other is primarily considered as a "locus in which speech is constituted", and Lacan frames this locus as the place of language, laws, norms, culture, and the Symbolic order itself (Lacan et al., 1993). Now we could understand the Other in the context of FEP. Firstly, all of the above Symbolic structures emerge from collective behaviors of human society, it is a collective structure that is contributed, acknowledged, and followed by most of the members. In terms of language, the Other is the 'treasure of signifiers', and as for norms, the Other is a mundane life routine for everyone. Imagining the Other as an average person, a compression of collective dynamics, and an agent B in Figure 4. **a**, many complex interpretations of Lacan turn obvious. For example, the desire to specific person (agent C) is the desire to the Other (agent B) by nature. And internalization of the Other, or entering the Symbolic order, is actually letting agent B occupy some parts of agent A, therefore, is a radical alienation (Lacan, 1988).

This part progressively explores the multi-layered manifestations of psychological dynamics through the construction and simulation of computational models based computational Lacanian psychoanalysis. It examines the tripartite structure of an individual subject, the dynamics of desire and partial synchronization between two subjects, and extends to the collective behaviors of multiple subjects involving the concept of the Other. With more fined engineering, we expect a digital twin mind that reproduces one's cognitive pattern, interpersonal relations, and social milieu from the first principle.

**Concluding remarks**

In this paper, we have proposed computational Lacanian psychoanalysis for engineering digital twin mind. We seek to bridge cognitive computational neuroscience, psychoanalysis, and digital twin by employing free energy principle (FEP) as a unifying framework. By elucidating fundamental parallels between FEP and Lacanian psychoanalysis, we have introduced a novel conceptual model: the FEP-RSI model.

Our exploration began with individual cognitive modalities, which we aligned with specific brain regions and functions through an intuitive mapping. By constructing a recurrent generative model, we demonstrated how disturbances in one cognitive domain could propagate through the system, illustrating the interconnectivity and dynamic nature of mental processes. Moving from individual to interpersonal dynamics, we incorporated the concept of desire into our model, presenting it as a drive towards partial generalized synchronization within and between subjects. This approach allowed us to capture the nuanced interpersonal dynamics posited by Lacan, showing how desire and the symbolic interactions of individuals could lead to complex behavioral manifestations. Finally, we extended our model to consider collective dynamics, bringing multiple agents into a shared world. This extension aimed to simulate how individual desires and interactions contribute to the emergent properties of social groups and cultural constructs. The computational Lacanian psychoanalysis formalizes various Lacanian core concept, including logical time, lack, object petit a, desire, and the Other. And simulations reproduce Lacanian descriptive dynamics of psyche, helping us understand this extremely complex theory at the same time.

By leveraging computational tools and the robust theoretical foundation provided by FEP, we suggest a potential approach for future research that could further refine our understanding of mental processes. This integrated approach opens new avenues for the development of sophisticated digital twins of the human mind, capable of simulating individual and collective psychological dynamics. Further efforts will focus on a hierarchical and recurrent model that contains various cognitive modalities, aiming to refine the precision and granularity with which these modalities interact and influence one another. By constructing such 'cognition-omics', we envision developing a more robust and detailed digital twin that not only mimics the cognitive processes of an individual but also dynamically adapts to the changing psychological states and environmental inputs. And larger networks of multiple digital twins could provide valuable insights into social dynamics, cultural evolution, and the formation of collective beliefs and behaviors.

**References**

Albarracin, M., Bouchard-Joly, G., Sheikhbahaee, Z., Miller, M., Pitliya, R. J., & Poirier, P. (2024). Feeling our place in the world: an active inference account of self-esteem. *Neuroscience of Consciousness*, *2024*(1), niae007. https://doi.org/10.1093/nc/niae007

Blei, D. M., Kucukelbir, A., & McAuliffe, J. D. (2017). Variational Inference: A Review for Statisticians. *Journal of the American Statistical Association*, *112*(518), 859-877. https://doi.org/10.1080/01621459.2017.1285773

Butler, E. A., & Randall, A. K. (2013). Emotional Coregulation in Close Relationships. *Emotion Review*, *5*(2), 202-210.

https://doi.org/10.1177/1754073912451630

Champion, T., Grześ, M., & Bowman, H. (2021). Realizing Active Inference in Variational Message Passing: The Outcome-Blind Certainty Seeker. *Neural Computation*, *33*(10), 2762-2826. https://doi.org/10.1162/neco_a_01422

Dall'Aglio, J. (2019). Of brains and Borromean knots: A Lacanian meta-neuropsychology. *Neuropsychoanalysis*, *21*(1), 23-38.

Durán, J. M. (2020). What is a Simulation Model? *Minds and Machines*, *30*(3), 301-323. https://doi.org/10.1007/s11023-020-09520-z

Evans, D. (2006). *An introductory dictionary of Lacanian psychoanalysis*. Routledge.

Feldstein, R., Fink, B., & Jaanus, M. (1995). *Reading Seminar XI: Lacan's Four Fundamental Concepts of Psychoanalysis: The Paris Seminars in English*. State University of New York Press.

Fink, B. (2017). *Lacan on love: An exploration of Lacan's seminar VIII, transference*. John Wiley & Sons.

Friston, K. J., & Frith, C. D. (2015). Active inference, communication and hermeneutics. *Cortex*, *68*, 129-143. https://doi.org/https://doi.org/10.1016/j.cortex.2015.03.025

Gardner, M. P. H., Schoenbaum, G., & Gershman, S. J. (2018). Rethinking dopamine as generalized prediction error. *Proceedings of the Royal Society B: Biological Sciences*, *285*(1891), 20181645. https://doi.org/doi:10.1098/rspb.2018.1645

Hasler, M., Maistrenko, Y., & Popovych, O. (1998). Simple example of partial synchronization of chaotic systems. *Physical Review E*, *58*(5), 6843-6846. https://doi.org/10.1103/PhysRevE.58.6843

Heins, C., Millidge, B., da Costa, L., Mann, R., Friston, K., & Couzin, I. (2023). Collective behavior from surprise minimization. arXiv:2307.14804. Retrieved July 01, 2023, from https://ui.adsabs.harvard.edu/abs/2023arXiv230714804H

Heins, R., Millidge, B., Demekas, D., Klein, B., Friston, K., Couzin, I., & Tschantz, A. (2022). pymdp: A Python library for active inference in discrete state spaces. *Journal of Open Source Software*, *7*, 4098. https://doi.org/10.21105/joss.04098

Hipólito, I., & van Es, T. (2022). Enactive-dynamic social cognition and active inference. *Frontiers in Psychology*, *13*, 855074.

Hogendoorn, H. (2022). Perception in real-time: predicting the present, reconstructing the past. *Trends in Cognitive Sciences*, *26*(2), 128-141. https://doi.org/https://doi.org/10.1016/j.tics.2021.11.003

Hook, D., Neill, C., & Vanheule, S. (2022). *Reading Lacan's Écrits: From 'Logical Time'to 'Response to Jean Hyppolite'*. Routledge.

Hook, D., & Vanheule, S. (2022). *Lacan on depression and melancholia*. Taylor & Francis.

Isomura, T., Kotani, K., Jimbo, Y., & Friston, K. J. (2023). Experimental validation of the free-energy principle with in vitro neural networks. *Nature Communications*, *14*(1), 4547.

Jandackova, V. K., Paulik, K., & Steptoe, A. (2012). The impact of unemployment on heart rate variability: The evidence from the Czech Republic. *Biological Psychology*, *91*(2), 238-244. https://doi.org/https://doi.org/10.1016/j.biopsycho.2012.07.002

Kastel, N., & Hesp, C. (2021). Ideas worth spreading: a free energy proposal for cumulative cultural dynamics. Joint European Conference on Machine Learning and Knowledge Discovery in Databases,

Kastel, N., Hesp, C., Ridderinkhof, K. R., & Friston, K. J. (2023a). Small steps for mankind: Modeling the emergence of cumulative culture from joint active inference communication. *Frontiers in neurorobotics*, *16*, 944986.

Kastel, N., Hesp, C., Ridderinkhof, K. R., & Friston, K. J. (2023b). Small steps for mankind: Modeling the emergence of cumulative culture from joint active inference communication [Original Research]. *Frontiers in neurorobotics*, *16*. https://doi.org/10.3389/fnbot.2022.944986

Kriegeskorte, N., & Douglas, P. K. (2018). Cognitive computational neuroscience. *Nature Neuroscience*, *21*(9), 1148-1160.

https://doi.org/10.1038/s41593-018-0210-5

Lacan, J. (1988). The Ego in Freud's Theory and in the Technique of Psychoanalysis, 1954-1955. *(No Title)*.

Lacan, J. (2001). *Ecrits: A selection*. Routledge.

Lacan, J., Miller, J.-A. E., & Grigg, R. T. (1993). The seminar of Jacques Lacan, Book 3: The psychoses 1955–1956. Translation of the seminar that Lacan delivered to the Société Française de Psychoanalyse over the course of the academic year 1955–1956.,

Lanillos, P., & Cheng, G. (2020). Robot Self/Other Distinction: Active Inference Meets Neural Networks Learning in a Mirror. In *ECAI 2020* (pp. 2410-2416). IOS Press.

Mills, J., & Downing, D. L. (2018). *Lacan on Psychosis: From Theory to Praxis*. Routledge.

Miyamoto, K., Rushworth, M. F. S., & Shea, N. (2023). Imagining the future self through thought experiments. *Trends in Cognitive Sciences*, *27*(5), 446-455. https://doi.org/10.1016/j.tics.2023.01.005

Mooij, A. (2012). *Psychiatry as a human science: Phenomenological, hermeneutical and Lacanian perspectives* (Vol. 18). Rodopi.

Ogolsky, B. G., Mejia, S. T., Chronopoulou, A., Dobson, K., Maniotes, C. R., Rice, T. M., Hu, Y., Theisen, J. C., & Carvalho Manhães Leite, C. (2022). Spatial proximity as a behavioral marker of relationship dynamics in older adult couples. *Journal of Social and Personal Relationships*, *39*(10), 3116-3132. https://doi.org/10.1177/02654075211050073

Ormrod, J. S. (2014). Fantasy in Lacanian Theory. In J. S. Ormrod (Ed.), *Fantasy and Social Movements* (pp. 98-133). Palgrave Macmillan UK. https://doi.org/10.1057/9781137348173_4

Palumbo, R. V., Marraccini, M. E., Weyandt, L. L., Wilder-Smith, O., McGee, H. A., Liu, S., & Goodwin, M. S. (2017). Interpersonal Autonomic Physiology: A Systematic Review of the Literature. *Personality and Social Psychology Review*, *21*(2), 99-141. https://doi.org/10.1177/1088868316628405

Pan, Y., Wen, Y., Jin, J., & Chen, J. (2023). The interpersonal computational psychiatry of social coordination in schizophrenia. *The Lancet Psychiatry*. https://doi.org/10.1016/S2215-0366(23)00146-3

Parr, T., & Pezzulo, G. (2021). Understanding, Explanation, and Active Inference [Original Research]. *Frontiers in Systems Neuroscience*, *15*. https://doi.org/10.3389/fnsys.2021.772641

Parr, T., Pezzulo, G., & Friston, K. J. (2022). *Active inference: the free energy principle in mind, brain, and behavior*. MIT Press.

Paulus, M. P., Feinstein, J. S., & Khalsa, S. S. (2019). An active inference approach to interoceptive psychopathology. *Annual review of clinical psychology*, *15*, 97-122.

Piekarski, M. (2023). *Incorporating (variational) free energy models into mechanisms (Preprint. Forthcoming in Synthese)*.

Priorelli, M., & Stoianov, I. P. (2023). Flexible intentions: An Active Inference theory [Original Research]. *Frontiers in Computational Neuroscience*, *17*. https://doi.org/10.3389/fncom.2023.1128694

Ramstead, M. J. D., Hesp, C., Tschantz, A., Smith, R., Constant, A., & Friston, K. (2021). Neural and phenotypic representation under the free-energy principle. *Neuroscience & Biobehavioral Reviews*, *120*, 109-122. https://doi.org/https://doi.org/10.1016/j.neubiorev.2020.11.024

Rescorla, M. (2021). Bayesian modeling of the mind: From norms to neurons. *WIREs Cognitive Science*, *12*(1), e1540. https://doi.org/https://doi.org/10.1002/wcs.1540

Riemer, M. (2018). Delusions of control in schizophrenia: Resistant to the mind's best trick? *Schizophrenia Research*, *197*, 98-103. https://doi.org/https://doi.org/10.1016/j.schres.2017.11.032

Sass, L. A. (2015). Lacan: the mind of the modernist. *Continental Philosophy Review*, *48*(4), 409-443. https://doi.org/10.1007/s11007-015-9348-y

Schoeller, F. (2019). Introduction to the special issue on physics of mind. *Physics of Life Reviews*, *31*, 1-10.

https://doi.org/https://doi.org/10.1016/j.plrev.2019.11.007

Sedley, W., Kumar, S., Jones, S., Levy, A., Friston, K., Griffiths, T., & Goldsmith, P. (2024). Migraine as an allostatic reset triggered by unresolved interoceptive prediction errors. *Neuroscience & Biobehavioral Reviews*, 105536.

Sikora, G. (2022). An economic model of the drives from Friston's free energy perspective [Hypothesis and Theory]. *Frontiers in Human Neuroscience*, *16*. https://doi.org/10.3389/fnhum.2022.955903

Smith, R., Parr, T., & Friston, K. J. (2019). Simulating emotions: An active inference model of emotional state inference and emotion concept learning. *Frontiers in Psychology*, *10*, 489395.

Solms, M. (2019). The Hard Problem of Consciousness and the Free Energy Principle [Hypothesis and Theory]. *Frontiers in Psychology*, *9*. https://doi.org/10.3389/fpsyg.2018.02714

Tao, F., Xiao, B., Qi, Q., Cheng, J., & Ji, P. (2022). Digital twin modeling. *Journal of Manufacturing Systems*, *64*, 372-389. https://doi.org/https://doi.org/10.1016/j.jmsy.2022.06.015

Tremblay, S. C., Essafi Tremblay, S., & Poirier, P. (2021). From filters to fillers: an active inference approach to body image distortion in the selfie era. *AI & society*, *36*, 33-48.

Vasil, J., Badcock, P. B., Constant, A., Friston, K., & Ramstead, M. J. D. (2020). A World Unto Itself: Human Communication as Active Inference [Hypothesis and Theory]. *Frontiers in Psychology*, *11*. https://doi.org/10.3389/fpsyg.2020.00417

Vilas, M. G., Auksztulewicz, R., & Melloni, L. (2022). Active inference as a computational framework for consciousness. *Review of Philosophy and Psychology*, *13*(4), 859-878.